\documentclass[]{spie}

\usepackage{xspace}
\usepackage[utf8]{inputenc} 
\usepackage{array}
\usepackage{amsmath,amsfonts,amssymb,textcomp}
\usepackage{mathtools}
\usepackage{mathrsfs}
\usepackage{graphicx}
\graphicspath{{figs/},{../2016-05-23/}}

\makeatletter
\newcommand{\MathFuncName}[1]{{\operator@font #1}}
\newcommand{\MathFunc}[1]{\mathop{\operator@font #1}\nolimits}
\newcommand{\MathFuncWithLimits}[1]{\mathop{\operator@font #1}\limits}
\makeatother

\newcommand{\degree}{{\text{\textdegree}}}

\newcommand{\V}[1]{\boldsymbol{#1}}      
\newcommand{\M}[1]{\mathbf{#1}}          

\newcommand{\diag}{\MathFunc{diag}}      
\newcommand{\prox}{\operatorname{prox}}  

\newcommand{\Fig}[1]{Figure~\ref{#1}}

\newcommand{\YORICK}[1]{}

\newcommand{\Norm}[1]{\left\Vert #1\right\Vert}

\newcommand{\Abs}[1]{\left\vert #1\right\vert}

\newcommand{\argmin}{\MathFuncWithLimits{arg\,min}}

\newcommand{\Reals}{\mathbb{R}}

\newcommand{\Complexes}{\mathbb{C}}

\newcommand{\eg}{\emph{e.g.}\xspace}
\newcommand{\ie}{\emph{i.e.}\xspace}
\newcommand{\etal}{\emph{et al.}\xspace}

\newcommand{\micron}{\text{\textmu{}m}} 

\usepackage[utf8]{inputenc}
 
\title{Back-propagating the light of field stars to probe telescope mirrors
aberrations}

 \author{F. Soulez\supit{a},  F. Courbin\supit{b} and M. Unser\supit{a}
 \skiplinehalf
           \supit{a}Biomedical Imaging Group, \'Ecole polytechnique fédérale de
           Lausanne
(EPFL),  CH-1015 Lausanne, Switzerland.\\
            \supit{b} Laboratoire d'Astrophysique,
\'Ecole polytechnique fédérale de Lausanne (EPFL) Observatoire de Sauverny
CH-1290 Versoix, Switzerland
                   }

\begin{document}
\maketitle

\begin{abstract} 
We propose a
wavefront-based method to estimate the PSF over the whole field of view. This
method estimate the aberrations of all the mirrors of the telescope using only
field stars. In this proof of concept paper, we described the method and
present some qualitative results.
\end{abstract}

\section{Motivation} 
The Euclid \cite{LaureijsDuvetEscuderoSanzEtAl2010} and WFIRST missions
\cite{SpergelGehrelsBreckinridgeEtAl2013} will probe dark matter distribution using weak
gravitational lensing.
The precision needed on galaxy shape measurements  required by  weak lensing
imposes stringent requirements on the PSF knowledge. The anisoplanatism of such 
wide-field telescope   can not be neglected and the PSF have to be
estimated for every position in the field of view.  Field stars can
give PSF measurements at random positions across the field of view.
However, for weak lensing, PSF must be computed at each galaxy position
(\ie between  field stars). The problem is thus twofold:
\begin{itemize}
  \item \emph{PSF estimation} at the position of each field star  from its
  noisy observations,
  \item \emph{PSF interpolation} at each galaxy position. 
\end{itemize}
There are mainly two approaches to solve the PSF estimation problem: (i)   image
domain methods that parameterize PSF with
pixels\cite{MboulaStarckRonayetteEtAl2015} and (ii)  pupil domain
methods. In the latter case, the PSF is described  as a function of
aberrations in the entrance pupil of the telescope. These pupil estimation
methods relies on  phase retrieval algorithms and most of it were conceived to
estimate Hubble Space Telescope aberrations at the beginning of the 90's
\cite{FienupMarronSchulzEtAl1993,Fienup1999,RoddierRoddier1993,ReddingDumontYu1993,lyon1997hubble,KristBurrows1995}.
The interpolation problem is then solved using a model of pupil aberration
variation across the field of view.

In this paper,  we propose to solve both problems jointly using a
wavefront based method to estimate the PSF over the whole field of view. 
Indeed, the PSFs at every position of the field of view are fully characterized 
by aberrations of each optical surface of the telescope and can be computed
using  Fourier optics  propagation.
Although these aberrations can be calibrated on ground, it is probable that they
will not remain stable enough after launch. One possible way to measure the
wavefront on orbit would be to strongly defocus and refocus the telescope, an 
operation that is  risky and therefore highly unlikely to be implemented by
space agencies.

In this proof of concept paper, we  propose a method to use scientific
observations to estimate wavefront aberrations on the few optical surfaces of a
space  telescope. It uses each observed bright star as a source of a coherent
plane wave to probe these aberrations as done for
diffraction tomography \cite{KamilovPapadopoulosShorehEtAl2015}. This method can
monitor the surface of every telescope mirrors bringing a new access to all its
optical component status without any need to move optical elements. In addition,
as it  use stars present in the scientific channel, it does not require any
additional calibration time. Finally, the knowledge of these optical surfaces
will give the mean to estimate the PSF at all wavelengths and in each point of
the field of view solving the problem of PSF interpolation on positions of
lensed galaxies.

Determining mirrors aberrations using many images of stars is
solved in an inverse problem framework.  For each star, the forward model
consists of free space propagation of a plane wave (whose angle is given by the
star position) across the telescope optics ended by intensity recording in the
detector plane. This model is non linear, however, as propagation between each
mirror is a linear operation, modeling errors can be back-propagated and used to
update the estimated aberrations of each mirrors. These back-propagated errors
for many stars across the field of view are used by a continuous optimization
algorithm (VMLMB, \cite{Thiebaut2002}) to probe
precisely these aberrations. In this algorithm, the phase retrieval problem given the
measured intensity is solved by the mean of an adapted  proximity operator 
\cite{Schutz2014PAINTER}.

\section{Image formation model}
\label{sec:ForwardModel}
The forward model links the incoming wave $w_1(x,y)$ arriving on the telescope
and the image recorded on the detector given the telescope parameters and its
aberrations $\V{\alpha}$. This model has two main parts:
\begin{itemize}
  \item the propagation of the incoming wave $w_1(x,y)$ through the telescope to
  the detector plane,
  \item the measurement by the detector which records only the intensity (\ie
  the squared modulus of the complex amplitude of the light in the detector
  plane) and is plagued by measurement noises such as both photon noise and read
  out noise.
\end{itemize}

\subsection{The telescope model}

 The incoming wavefront emitted at wavelength
$\lambda$ by a single star at angular position $(\theta_1, \theta_2)$
relatively to the telescope optical axis, can be modeled as a plane wave. Its
complex amplitude in the first mirror (M1) plane is given by
\begin{equation}\label{eq:w1}
w_1(x,y) =  \exp\left(\imath \left( x\,\sin(\theta_1) / \lambda + y
\,\sin(\theta_2 )/ \lambda \right) \right)\,.
\end{equation} 

To define our forward model, this wave is adequately sampled on $N$ pixels and
we adopt a vector representation:  $\V{w}_1 = (w_{1,1},\dots, w_{1,N})$.
The propagation of this wave through the telescope can be decomposed as a
sequence of $K$ similar operations, where $K$ is the number of optical
interfaces (mirrors, lenses and the detector). For each interface $k$, the
incoming wave (\ie the wave right before the interaction with the interface) can 
be itself modeled as a sequence of linear operations
\begin{equation} \label{eq:TelescopModel}
\V{w}_k(\V{\alpha}) = \M{H}_{k-1}\,\M{M}_{k-1}\,\M{A}_{k-1}(\V{\alpha}_k)
\,\V{w}_{k-1}(\V{\alpha})
\end{equation}
where $\M{M}_{k-1}$ and $\M{A}_{k-1}(\V{\alpha}_k)$  are two diagonal operators
accounting for the effect of the $(k-1)^{\mathrm{th}}$ interface and its aberration
respectively and
$\M{H}_{k-1}$ is a propagation operator from the
interface $k-1$ to the interface $k$.
 All these operators are square matrices in $
\Complexes^{N\times N}$. The aberration operator $\M{A}_{k-1}(\V{\alpha}_k)$ is
a function of the unknown aberration  parameters $\V{\alpha}_k$ that will be
estimated in our methods. Whereas the model described in Equation
(\ref{eq:TelescopModel}) is linear in $\V{w}$, it is highly non-linear in
$\V{\alpha}$.

\subsubsection{Mirrors}
A mirror $M_k$ modifies the incoming wave in two ways; (i) it cuts the light
outside of its support $p_k(x,y)$ and (ii)  it  adds a space-varying phase term:
\begin{equation}
m_k(x,y) = p_k(x,y) \exp\left(\imath\,2\,\phi_k(x,y) \right)\,
\end{equation}
where $\phi_k(x,y)$ is the sagitta of the mirror defined by 
\begin{equation}
\phi_k(x,y) = \frac{x^2 + y^2}{R_k + \sqrt{R_k^2 - (1-\epsilon^2_k)\,(x^2 +
y^2)}}\,,
\end{equation}
where $R_k$ and $\epsilon_k$ are the radius of curvature and the eccentricity respectively.

The mirror support $p_k$ is defined by
\begin{equation}
 p_k(x,y)  = \left\{
\begin{array}{ll}
1\,, & \text{if } (x,y)  \text{ is inside the support}\\
0 \,, & \text{otherwise.}\\
\end{array}\right.
\end{equation} 

With an  adequate sampling of $m_k(x,y)$, the discrete operator $\M{M}_k$ is
diagonal and writes
\begin{equation}
\M{M}_k = \diag\left(\V{m}_k\right).
\end{equation}

\subsubsection{Aberrations}

The aberrations are due to errors in the polishing of  mirrors  and optical
misalignement. They are described by additional phase terms in the plane of
each mirror. As the support of each mirror  is usually a disk,
the Zernike polynomials $\M{Z}$ provide a suitable basis to express these
aberrations.   The aberrations of the $k^\textrm{th}$
mirror  are described in the zernike basis $\M{Z}_k$ with the parameters 
$\V{\alpha}_k$. The aberration operator is 
\begin{equation}\label{eq:aberration}
\M{A}_k(\V{\alpha}_k) = \diag\left(\exp\left(\imath\, \M{Z}_k \,\V{\alpha}_k
\right)\right)\,.
\end{equation}

\subsubsection{Propagation}
The propagation operator $\M{H}_{k}$ from the interface
$k$ to the interface $k+1$ 
is modeled using the paraxial approximation. 
Given the size of most telescopes, the Fresnel number
is in general very high ($\propto 10^6$) and we define  the propagation operator 
$\M{H}_{k}$ using the angular spectrum method.

\subsection{Measurements}

The detector measures only the intensity of the light wave. The 
 forward model that links the complex amplitude in the detector plane $\V{w}_K
 \in \Complexes^N $ to  the measured  image intensities $\V{d}\in \Reals^{N}_{+} $
is then
\begin{equation}
d_n = \Abs{w_{K,n}}^2 + e_n\,,
\end{equation}
where $e_n$ is some measurement noise with spatially varying variance
$\sigma_n^2$ and $\Abs{w_{K,n}}^2$ denotes the squared modulus of $w_{K,n}$.

\section{Algorithm}

The goal of our algorithm is to estimate the vector of aberration parameters
$\V{\alpha}$ using observations of $S$ stars randomly distributed across the field of view. 
Assuming  Gaussian measurement noise $\V{e}$, the estimated aberration
parameters $\V{\alpha}^+ $ is the solution of the minimization problem:
\begin{equation}
\V{\alpha}^+ = \argmin_{\V{\alpha}} \sum^S_{s=1}  \sum^N_{n=1}
\frac{1}{\sigma_n^2} \left(\Abs{w^s_{K,n}(\V{\alpha})}^2 - d_n^s\right)^2 \,.
\end{equation}
$w^s_{K,n}$ is the complex amplitude at $n^{\text{th}}$ pixel of the detector
of the light emitted by the $s^{\text{th}}$ star. It is modeled using the
Equation~\ref{eq:TelescopModel} with $\V{w}^s_1$ given by the
Equation~\ref{eq:w1}.

This problem can be
reformulated as a constrained problem:
\begin{equation}
 \V{\alpha}^+\argmin_{\V{\alpha}} \sum^S_{s=1}  \sum^N_{n=1}
\frac{1}{\sigma_n^2} \left(\Abs{y_n^s}^2 - d_n^s\right)^2 \,
\text{ subject to }	y_n^s = w^s_{K,n}(\V{\alpha}) \,,
\end{equation}
The Augmented  Lagrangian formulation of this constrained problem is:
\begin{equation}
\mathcal{L}(\V{\alpha},\V{t}, \V{u}) = \sum^S_{s=1}
 \sum^N_{n=1}
\frac{1}{\sigma_n^2} \left(\Abs{t^s_{n}}^2 - d_n^s\right)^2 +   \frac{\rho}{2}
\sum^S_{s=1} \Norm{w^s_K(\V{\alpha}) - \V{t}^s - \V{u}^s }_2^2\,,
\end{equation}
where $\V{u}^s$ are the  scaled Lagrange multipliers and $\rho>0$ is the
augmented penalty parameter.

Following Mourya \etal \cite{MouryaDenisBeckerEtAl2015}, we solve this problem
in a hierarchical way:

\begin{align}
 \V{\alpha} &= \argmin_{\V{\alpha}}\sum^S_{s=1}  \sum^N_{n=1} 
 \Norm{w^s_{K,n}(\V{\alpha}) - {t}_n^s(\V{\alpha}) - u_n^s }^2\label{eq:outer}
 \\
 \text{with } 
 t_n^s(\V{\alpha})  &= \argmin_{t \in \Complexes}
\frac{1}{\sigma_n^2}\left(\Abs{t}^2 - d_n^s\right)^2 +
\frac{\rho}{2} \Norm{t - w^s_{K,n}(\V{\alpha}) +
u_n^s }^2 \label{eq:inner}
 \end{align} 
 The Equation \ref{eq:outer} is solved using a continuous iterative
 optimization method (\eg quasi-Newton method). The inner
 Equation~\ref{eq:inner} is separable and consists on solving $S\times N$
 small 1D problems that can be easily parallelized. 
 At the end of each iteration $k$, we
 update the Lagrangian parameters:
 \begin{equation}
 \V{u}^{(k+1)} = \V{u}^{(k)} + \V{w}^{(k)} - \V{t}^{(k)}
 \end{equation}

\subsection{The phase retrieval problem}
The inner minimization problem in Equation \ref{eq:inner} is a phase retrieval
problem. It is separable and can  be defined as  the solution of the proximity operator of the function $f$:
\begin{eqnarray}
f(x) &=& \frac{1}{\sigma^2} (\Abs{x}^2 -
d)^2\,,\label{eq:Glkldef}\\
\label{eq:prox-def}
\prox_{1/\rho\,f}(t) &=& \argmin_{x \in \Complexes}
\left( \frac{1}{\rho}\, f(x) + \frac{1}{2}\,\Abs{x - t}^2\right)
\, .
\end{eqnarray}
This proximity operator has a closed form solution described in Schutz \etal
\cite{Schutz2014PAINTER}.

\subsection{The tomography problem}
The outer minimization in Equation \ref{eq:outer} is a tomography problem. It
can be rewritten as:
\begin{equation}
\V{\alpha} = \argmin_{\V{\alpha}} \sum^S_{s=1}
\Norm{\V{w}_{K}(\V{\alpha}) -
  \V{t} -  \V{u} }_2^2\label{eq:tomo}
\end{equation}
We solve this non linear problem using VMLMB \cite{Thiebaut2002}, a continuous
optimization routine. The needed derivatives are computed using the recursive
back-propagation algorithm described by Kamilov \etal
\cite{KamilovPapadopoulosShorehEtAl2015}.

\section{Results}
 \begin{table}
 \centering
 \begin{tabular}[h]{lc}
 \multicolumn{2}{c}{\textbf{Mirror 1}}\\
 Diameter (m) & $2.4$ \\
 Curvature radius (m)& $11.040$ \\
 Conic constant& $-1.0022985$\\
 \multicolumn{2}{c}{\textbf{Mirror 2}}\\
 Diameter (m) & $ 0.281$\\
 Curvature radius (m)& $-1.358$\\
 Conic constant&$-1.496$\\
 \multicolumn{2}{c}{\textbf{Distance (m)}}\\
 M1 to M2 & $4.9069$ \\
 M2 to detector & $6.4062$ \\
 \multicolumn{2}{c}{\textbf{Detector}  }\\
 Field of view (°) & $0.6 \times 0.6$ \\
 Pixel size ($\micron$)& $5$ \\
 Wavelength & $500\,$nm\\
 \end{tabular}
 \vspace{5mm}
 \caption{Telescope simulation parameters \label{tab:HST}}
 \end{table}
 
\begin{figure}[tb] \noindent\begin{minipage}[t]{.3\linewidth}
{\includegraphics[height=\linewidth,width=\linewidth]{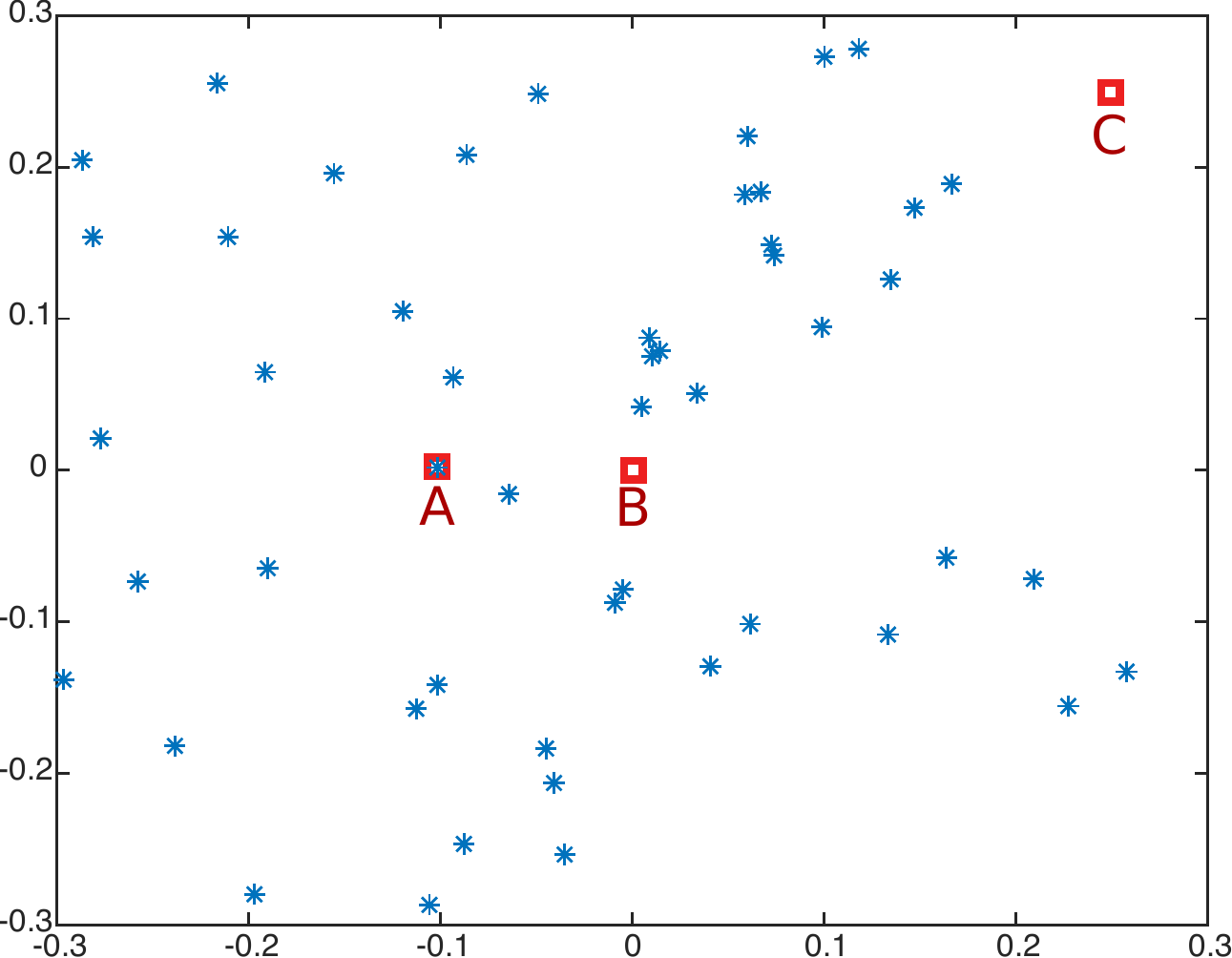}}
\caption{Distribution of the observed stars (blue stars) and test stars (red
square)  across the $0.6\,{\degree}\times0.6\,\degree$ field of
view.}
\label{fig:FoV}
\end{minipage}\hspace{1cm}
\begin{minipage}[t]{.3\linewidth}
{\includegraphics[height=\linewidth,width=\linewidth]{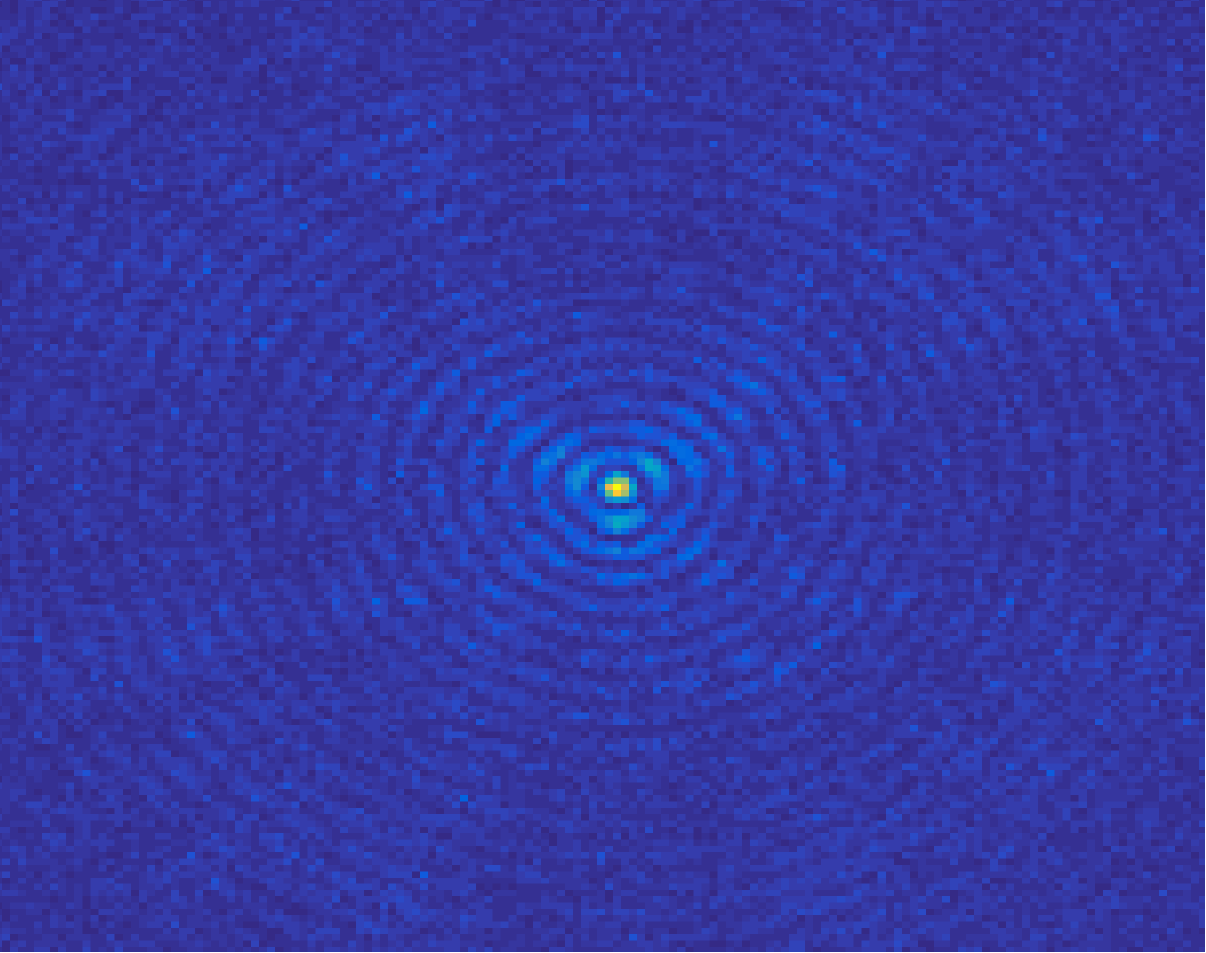}}
\caption{ $300\times300$ pixels central part of the recorded intensity for  the
star indicated by the letter A on
\Fig{fig:FoV}}
\label{fig:t50_noisy}
\end{minipage}
\end{figure}

\begin{figure*}
    \begin{minipage}[c]{\linewidth}
    \begin{small}
      \begin{tabular}[h]{m{1em} m{.3\columnwidth}  m{.3\columnwidth}
      m{.3\columnwidth}} & \centering Estimation &  \centering Truth &
      {\qquad \qquad No Aberration} \\
        A &
        \begin{minipage}[b]{.3\columnwidth} 
        \includegraphics[width=\columnwidth,height=\columnwidth]{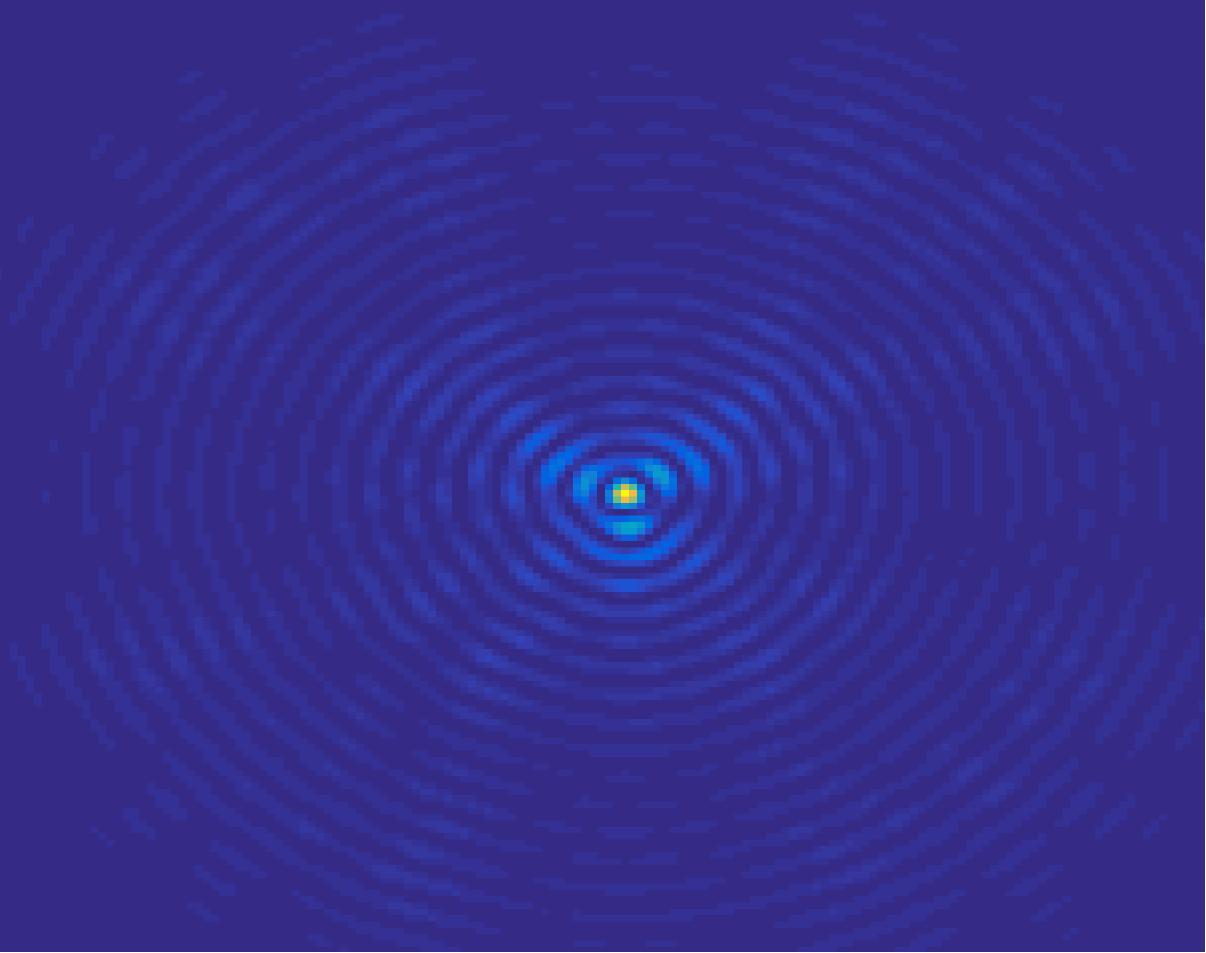}\end{minipage}
        &
        \begin{minipage}[b]{.3\columnwidth} 
        \includegraphics[width=\columnwidth,height=\columnwidth]{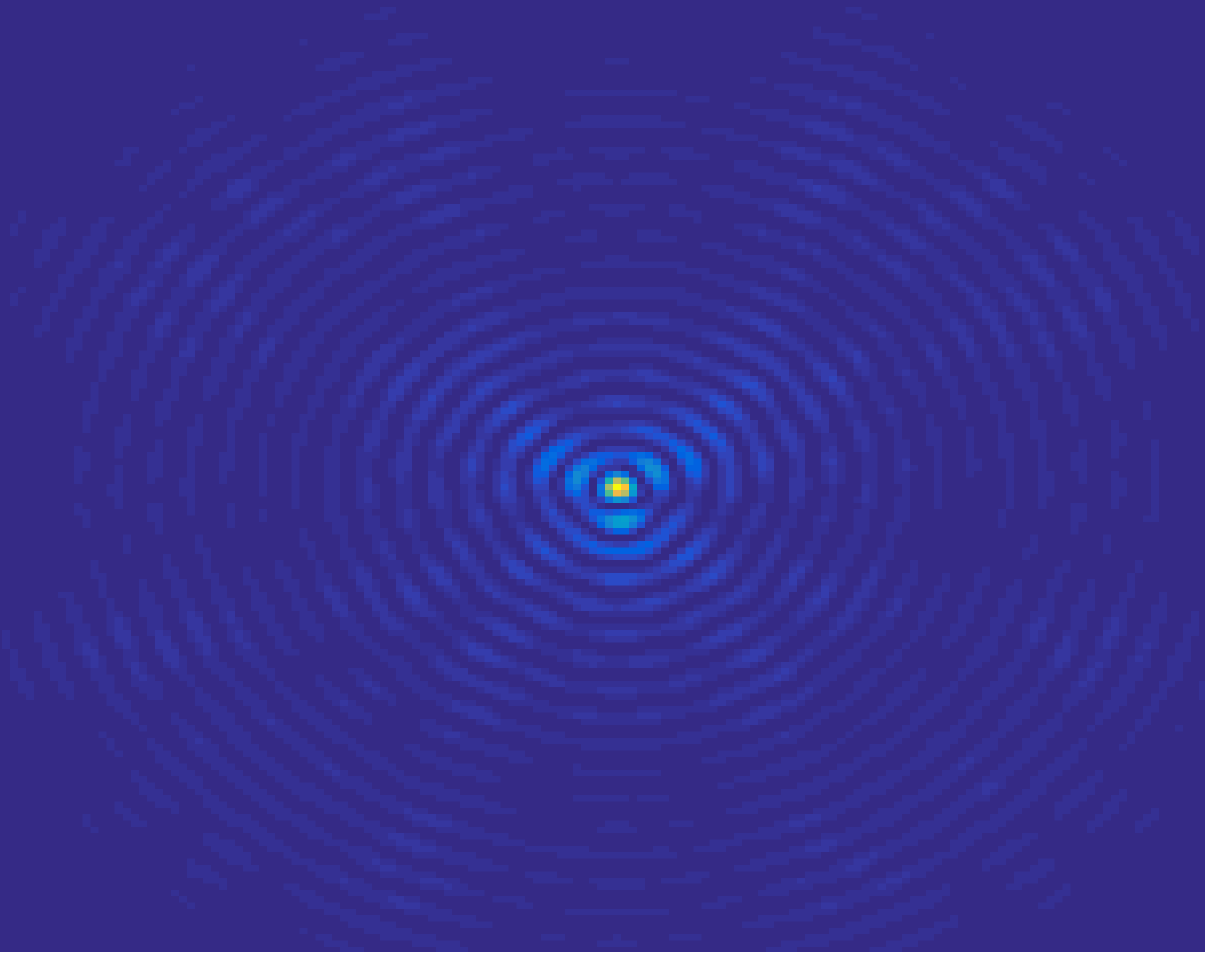}\end{minipage}
        &
        \begin{minipage}[b]{.3\columnwidth} 
        \includegraphics[width=\columnwidth,height=\columnwidth]{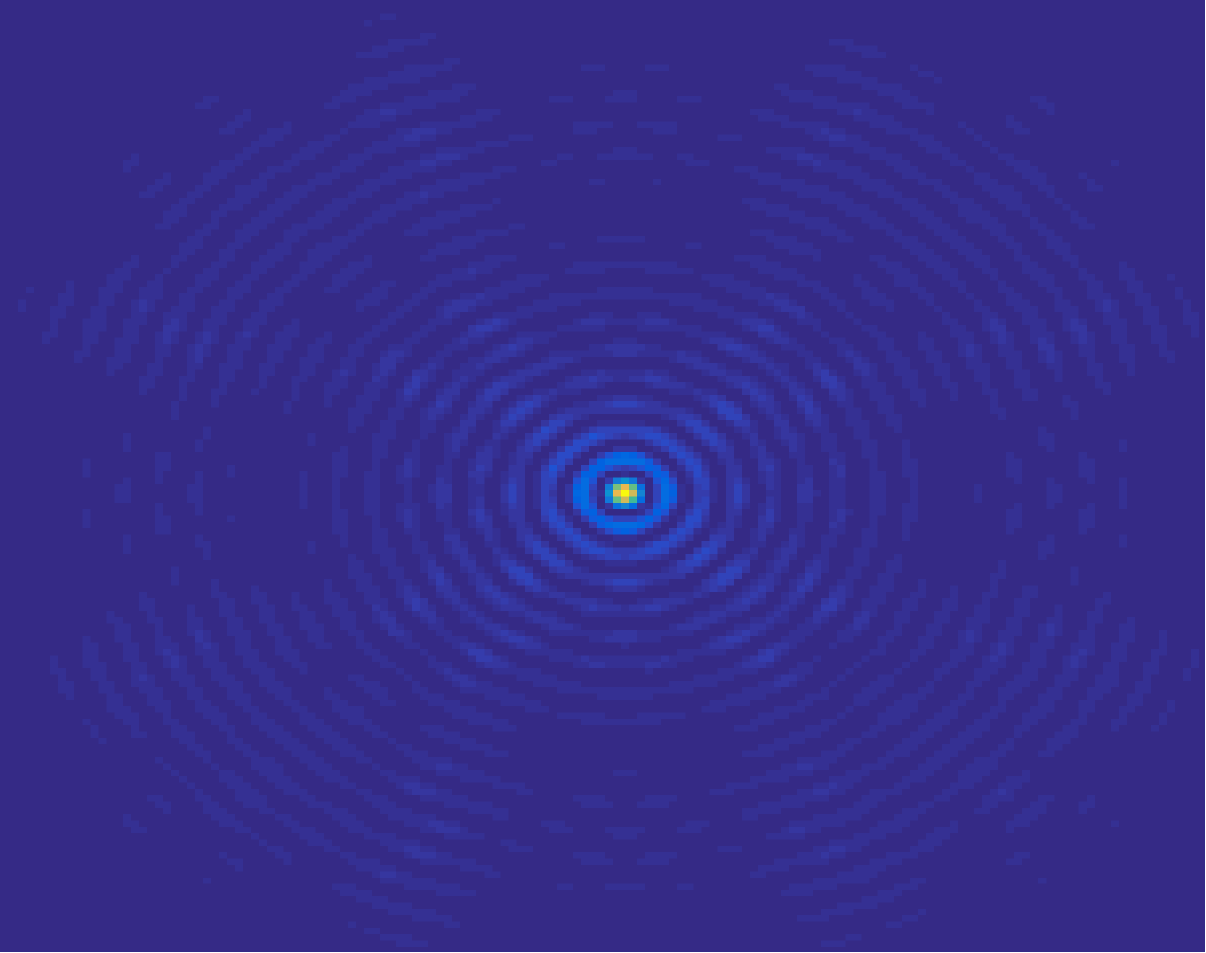} 
        \end{minipage} \\
        B&
        \begin{minipage}[b]{.3\columnwidth} 
        \includegraphics[width=\columnwidth,height=\columnwidth]{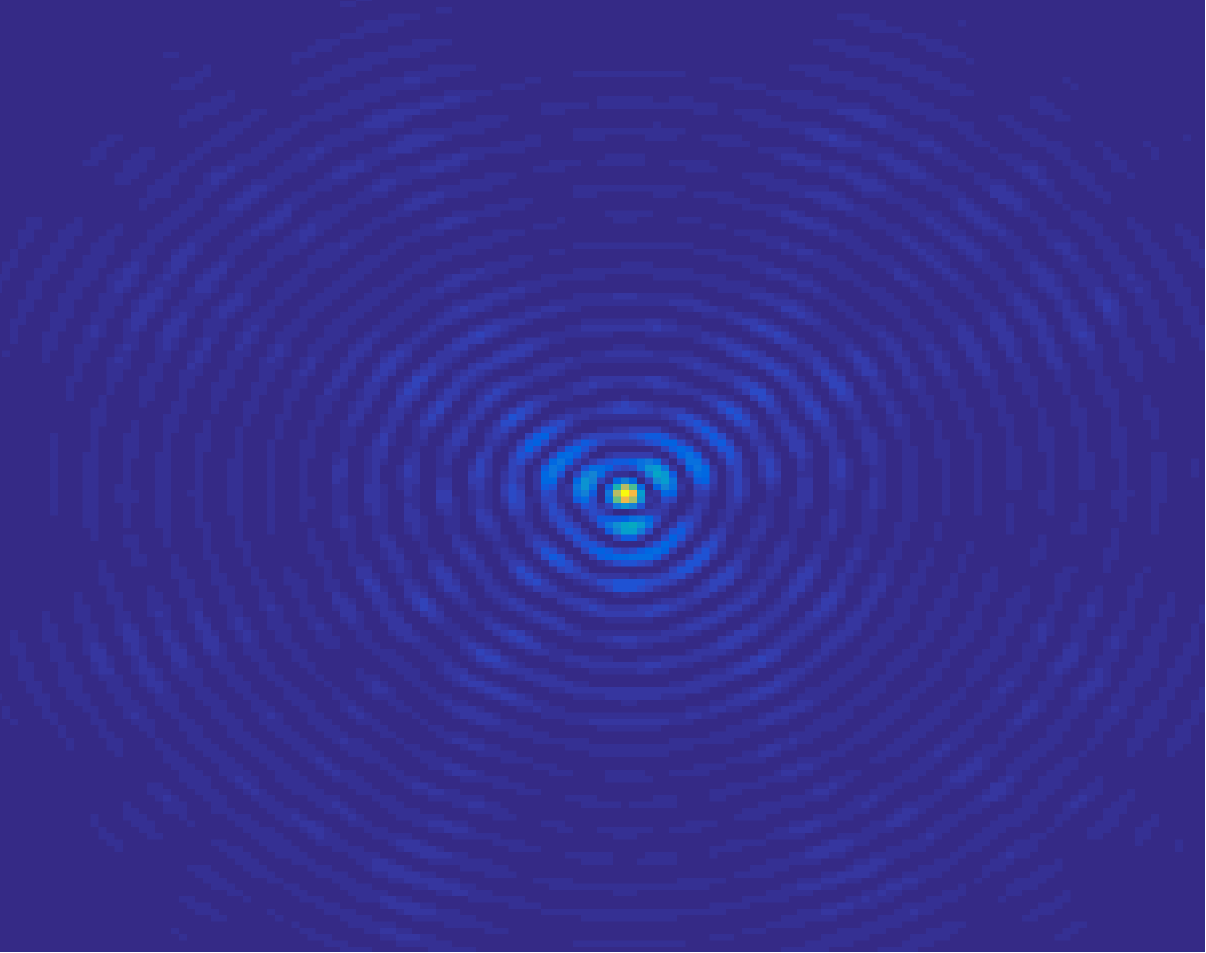}\end{minipage}
        &
        \begin{minipage}[b]{.3\columnwidth} 
        \includegraphics[width=\columnwidth,height=\columnwidth]{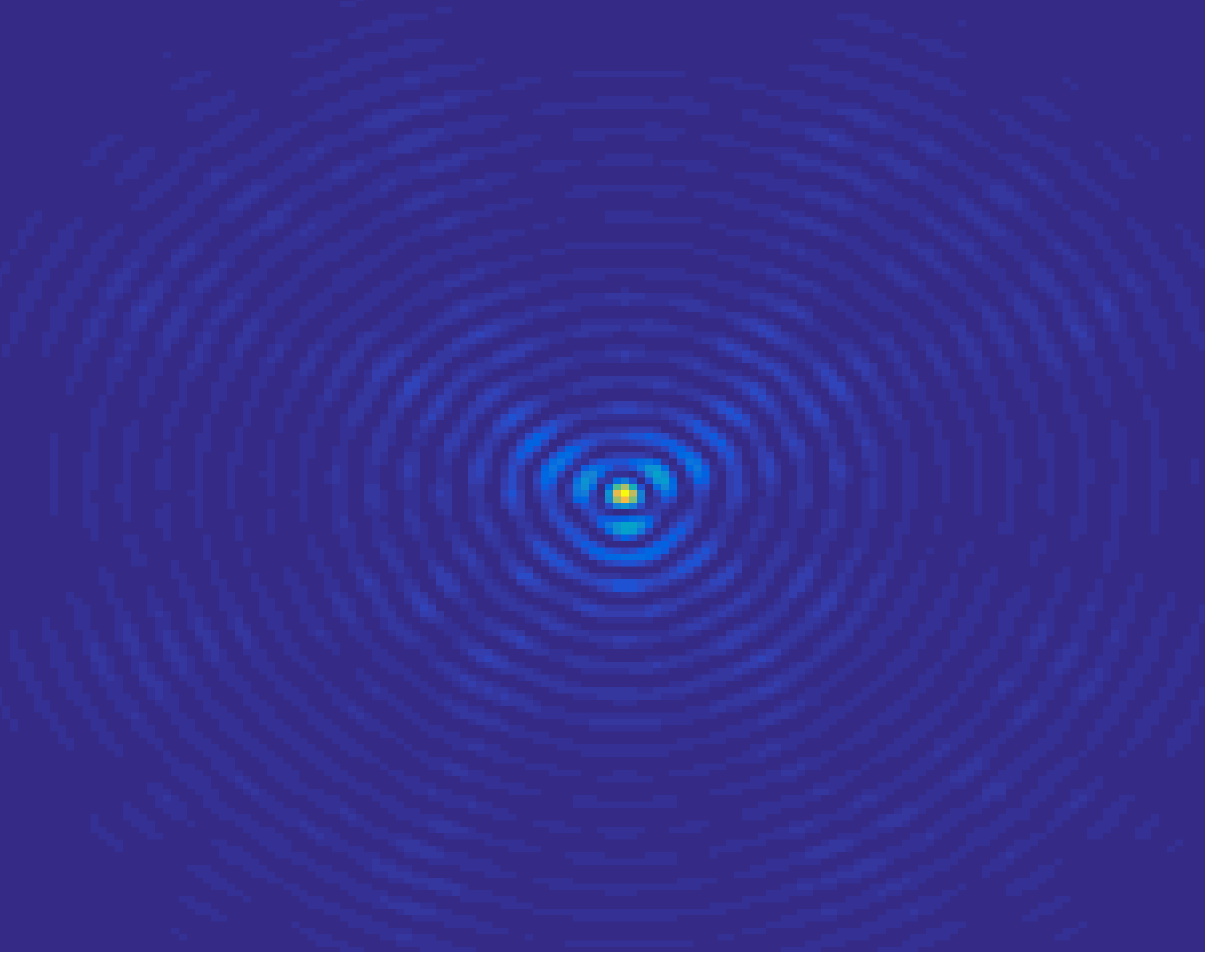}\end{minipage}
        &
        \begin{minipage}[b]{.3\columnwidth} 
        \includegraphics[width=\columnwidth,height=\columnwidth]{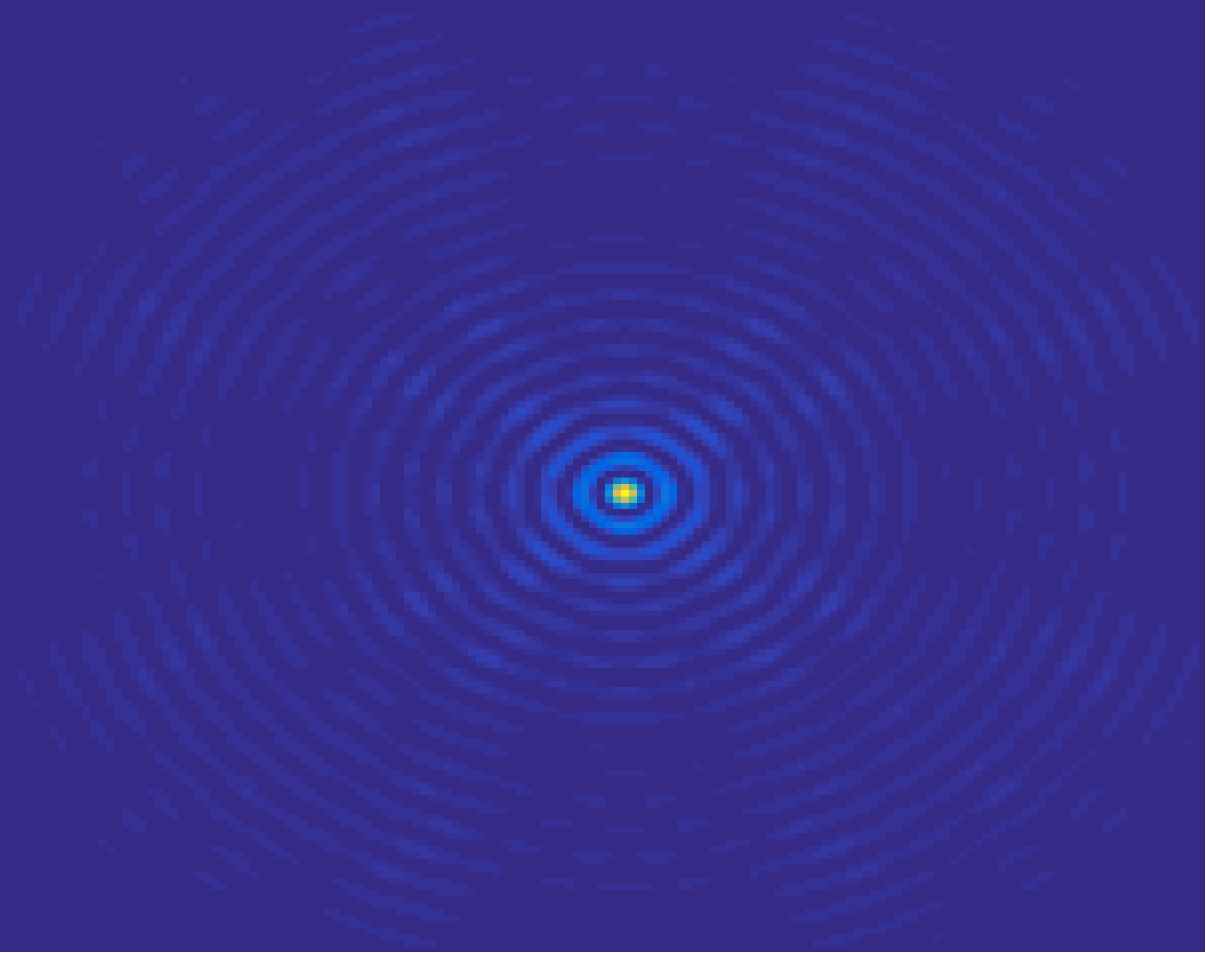} 
        \end{minipage} \\
        C &
        \begin{minipage}[b]{.3\columnwidth} 
        \includegraphics[width=\columnwidth,height=\columnwidth]{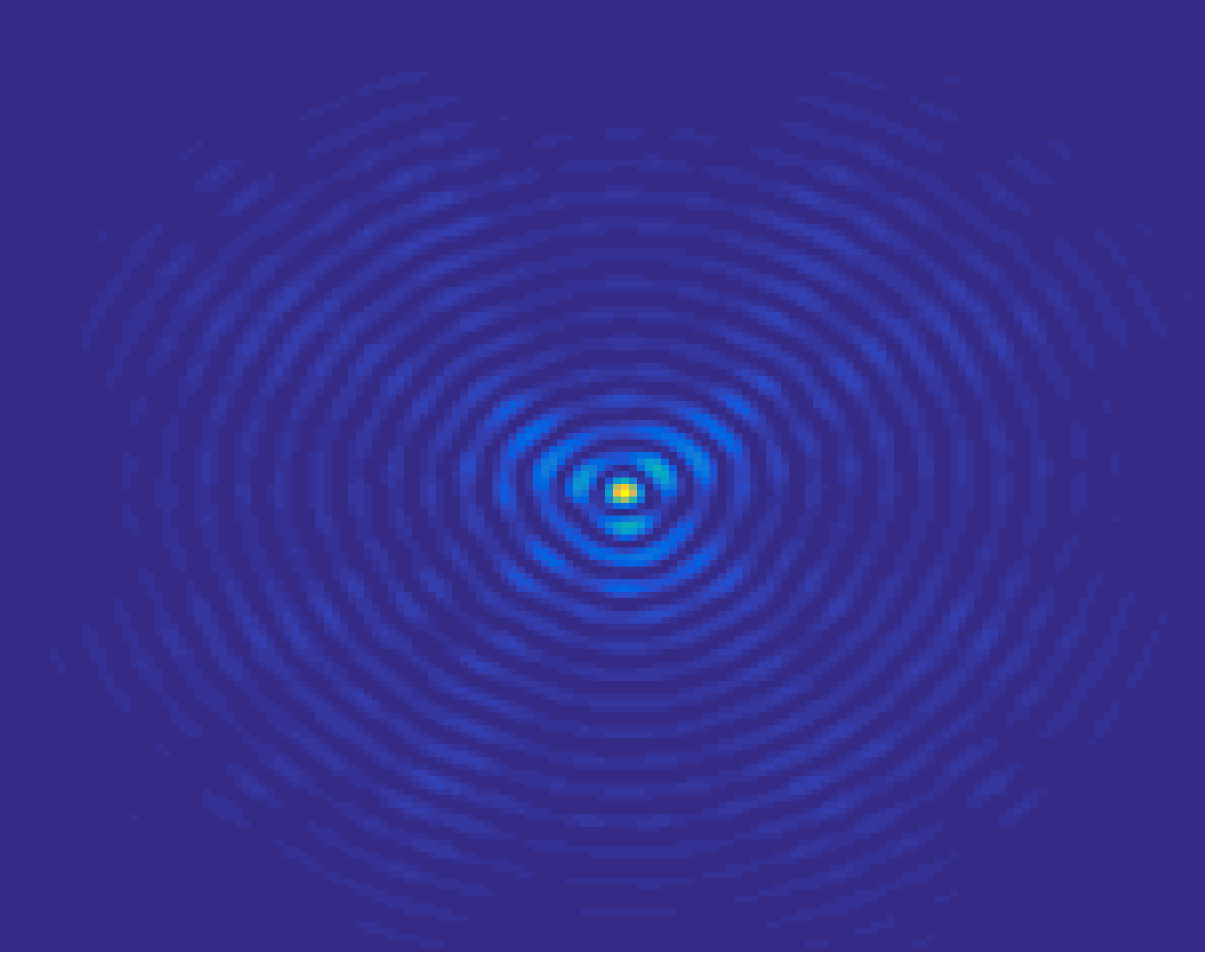}\end{minipage}
        &
        \begin{minipage}[b]{.3\columnwidth} 
        \includegraphics[width=\columnwidth,height=\columnwidth]{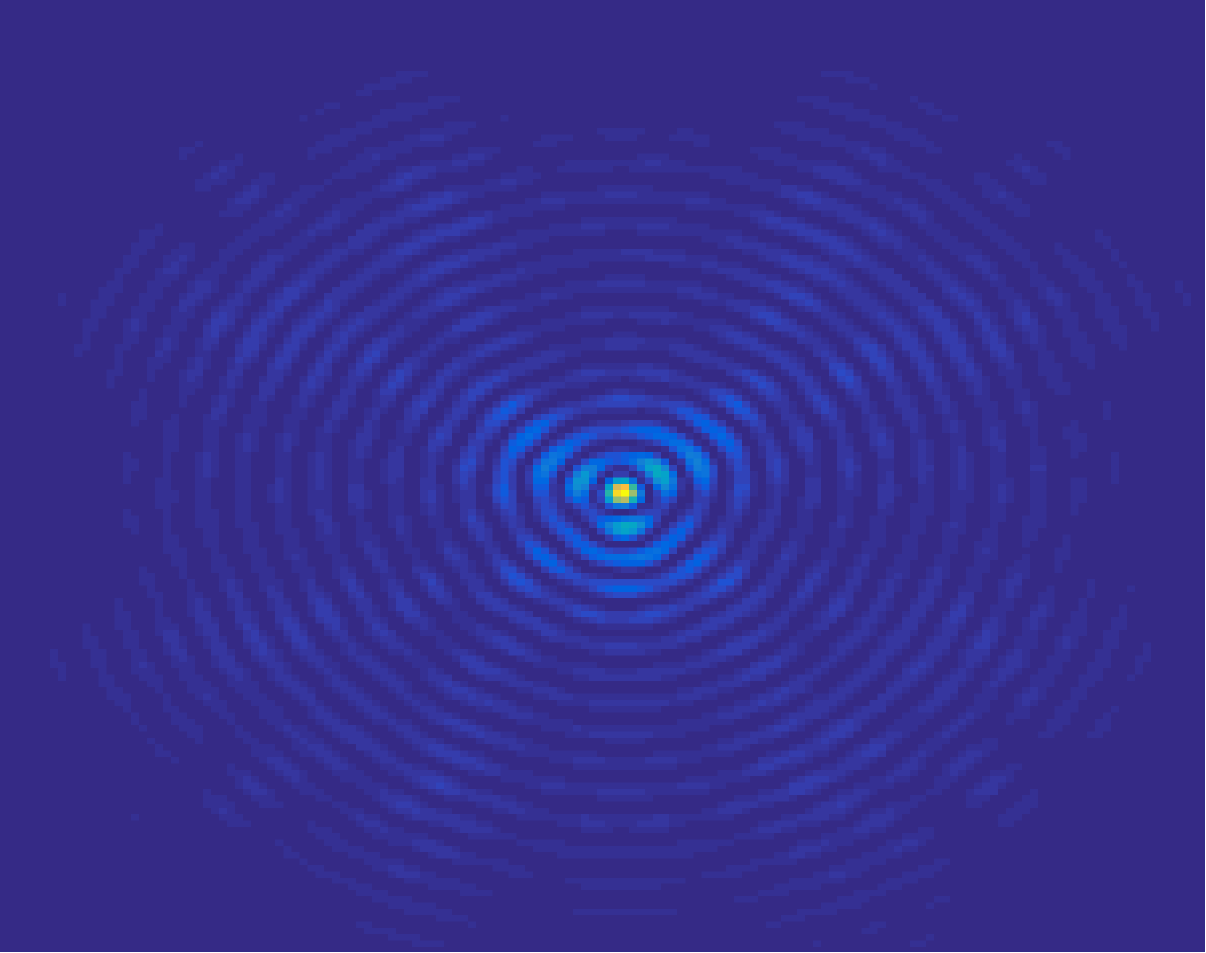}\end{minipage}
        &
        \begin{minipage}[b]{.3\columnwidth} 
        \includegraphics[width=\columnwidth,height=\columnwidth]{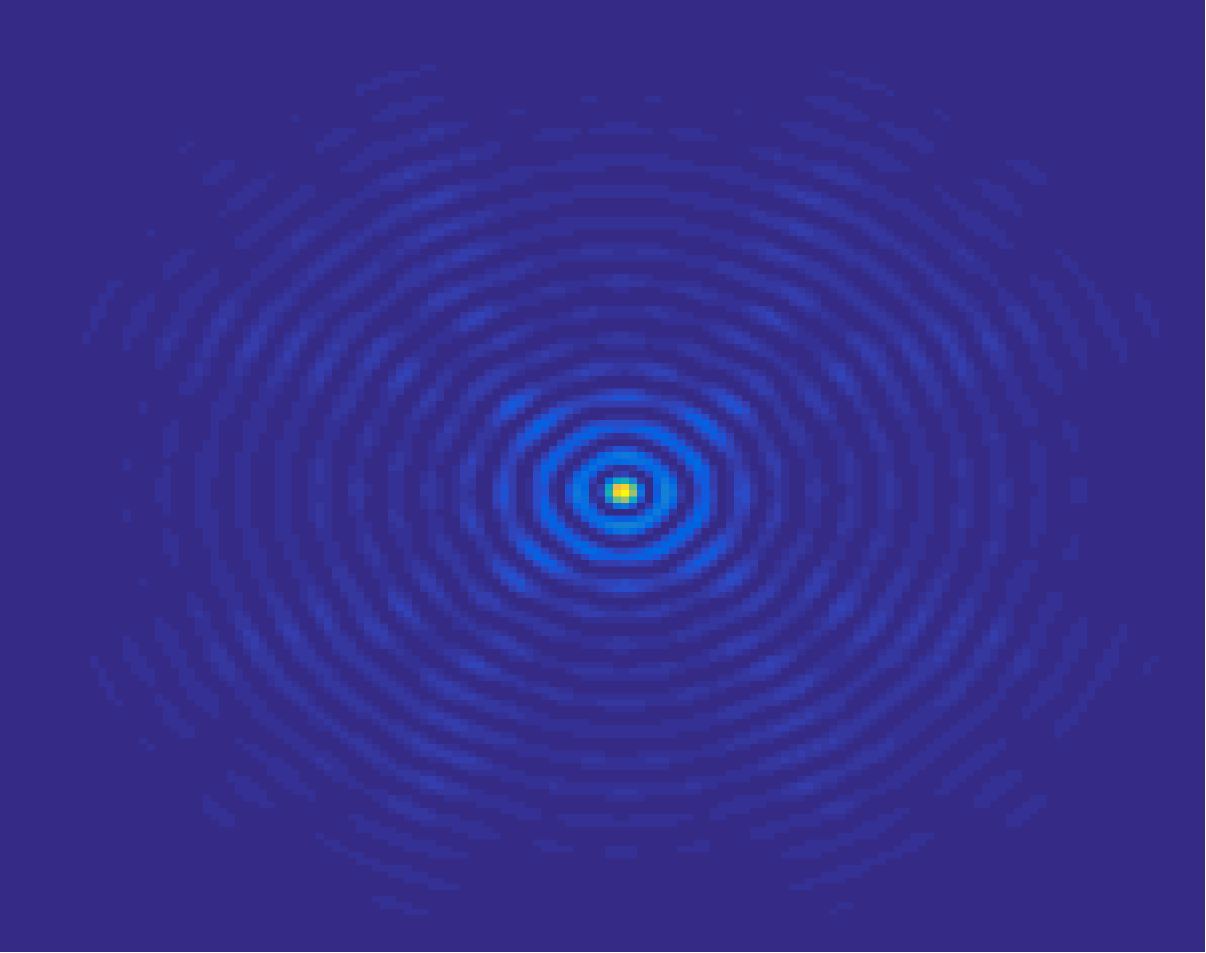} 
        \end{minipage} \\

      \end{tabular} 
    \end{small} 
    \caption{Estimation and  truth of image (PSF) of the star
    indicated by A, B and C on Figure~\ref{fig:FoV}. It is be compared to
    stars images for a perfect aberration-free telescope.}
    \label{fig:PSFs}
  \end{minipage}
\end{figure*}

 We have tested our algorithm on simulations. We have simulated a
 Richtey-Chrétien telescope similar to the Hubble Space Telescope. Its
 characteristics are given by the Table~\ref{tab:HST}. We introduce aberrations
 by drawing random coefficients of Zernike basis in
 Equation~\ref{eq:aberration}.  We used $56$ and $10$ coefficients for the
 aberration of the first and the second mirrors respectively.

 The dataset was generated with the telescope model described in
 Section~\ref{sec:ForwardModel}. 50 stars distributed randomly across the field
 of view were generated. Their positions are shown on Figure~\ref{fig:FoV}.
 Their fluxes were adjusted such that  $26000$ photons on average were
 recorded per star ; that corresponds to a maximum intensity of $256$ photons in
 the brightest pixel of the $1500\times1500$ pixels PSF. To simulate the
 detector, we add background noise of $5\,e^-$ and generate the data $\V{d}^s$ using the Poisson
 distribution $\mathcal{P}$:
 \begin{equation}
 d_n^s   = \mathcal{P}(\Abs{w^s_{K,n}}^2 + 5)\,.
 \end{equation}
The $300\times300$ pixels central part of the observation of the star indicated 
by an A on Figure~\ref{fig:FoV} is shown on Figure~\ref{fig:t50_noisy}.
 
We minimized Equation~\ref{eq:outer} using the
unaberrated telescope as a starting point $\V{\alpha} = \V{0}$.
To assess the performance of our method, we  simulate  observations of stars
that were not in the  data-set (denoted as B and C on Figure~\ref{fig:FoV})
using our aberrations estimate, the true aberrations and without any
aberrations. These stars B and C and one of the star  used in  aberrations
estimation (A) shown on Figure~\ref{fig:PSFs}. 
On this figure, we can see that, beginning from an aberration free model,
our algorithm successfully converges toward a PSF  very similar to the
ground truth.

\section{Conclusion and future works}
In this proof of concept paper, we show the validity of our approach to estimate
PSFs at every positions across the field of view without any need to carry out
in-flight telescope defocusing to measure directly the wavefront. It achieves to provide qualitatively good PSF even in noisy conditions and we are working on
quantitative results in term of ellipticiy and size of the estimated PSF. In
addition a lot of works has to be done to be able to process real data,
particularly it has to handle undersampled and broadband PSFs.

\section*{Acknowledgements}
This work is supported  by the Sinergia project “Euclid: precision cosmology in
the dark sector" from the Swiss National Science Foundation

\bibliographystyle{spiebib}
\bibliography{TelescopeTomography}

\end{document}